\documentclass{article}

\usepackage{arxiv}

\usepackage[utf8]{inputenc} 
\usepackage[T1]{fontenc}    
\usepackage{hyperref}       
\usepackage{url}            
\usepackage{booktabs}       
\usepackage{amsfonts}       
\usepackage{nicefrac}       
\usepackage{microtype}      
\usepackage{lipsum}		
\usepackage{graphicx}
\usepackage{doi}
\usepackage{amsmath} 
\usepackage{subcaption}

\title{Predictions in modified Glauber model of total charged-particle yields centrality dependence in O+O and Ne+Ne collisions at LHC}

\author{ \href{https://orcid.org/0009-0004-8531-5075}{\includegraphics[scale=0.06]{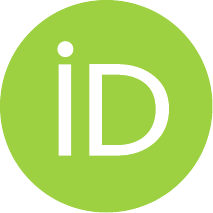}\hspace{1mm}Simak Svetlana}\thanks{
        \* Corresponding author} \\
	Laboratory of Ultra High Energy Physics\\
	Saint-Petersburg State University\\
	Saint-Petersburg, Russia \\
	\texttt{s.simak@spbu.ru} \\
	\And
	\href{https://orcid.org/0000-0003-3700-8623}{\includegraphics[scale=0.06]{orcid.pdf}\hspace{1mm}Feofilov Grigory} \\
	Laboratory of Ultra High Energy Physics\\
	Saint-Petersburg State University\\
	Saint-Petersburg, Russia \\
	\texttt{g.feofilov@spbu.ru} \\
}

\renewcommand{\shorttitle}{Predictions in MGM for O+O and Ne+Ne collisions at LHC }

\hypersetup{
pdftitle={A template for the arxiv style},
pdfsubject={q-bio.NC, q-bio.QM},
pdfauthor={David S.~Hippocampus, Elias D.~Striatum},
pdfkeywords={First keyword, Second keyword, More},
}

\begin{document}
\maketitle

\begin{abstract}
	In this article we present the results of application of the Monte Carlo modified Glauber model \cite{FI,FS,SF} for the predictions of collision centrality dependence of the total charged-particle yields for 16O +16O and 20Ne+20Ne colliding systems at the LHC.

Our model differs from the Standard Glauber model by the effective account of the energy losses in successive inelastic nucleon-nucleon collisions. For this purpose,  a single model parameter $k$ is defined  as a mean fraction of the momentum loss that happens in any binary nucleon collision  due to the production of multiple particles. Therefore, the decrease of the nucleon momentum after each inelastic collision leads to a corresponding reduction of the inelastic cross section and the mean multiplicity yield in the subsequent interaction. All these effects are taken into account in the MC model  
in each of the subsequent inelastic binary interactions.

We discuss the purely geometrical effects for these light colliding systems that could be considered useful in future studies of QGP properties in energy density scanning. Predictions in this single parameter model are based on the previous successful analysis~\cite{SF} of non-linear  centrality dependence of charged particle yields observed in Pb+Pb collisions by ALICE at the LHC.


\end{abstract}


\section{Introduction}

Studies of light-ion collisions at the LHC energies aim to explore the properties of hot,  strongly interacting matter created in case of small systems. For O+O and Ne+Ne interactions, of primary interest are the centrality dependence of charged-particle multiplicity, rapidity distributions, collective flow and parton energy losses.
The initial conditions in small systems are described, as in heavy-ion collisions, by centrality, linked to the impact parameter $b$ or the number of participating nucleons $N_{part}$. These quantities cannot be measured directly and are estimated through Glauber-type approaches.

In the Standard Glauber Model (SGM), the total charged-particle multiplicity $\langle N_{ch}^{tot} \rangle$ is initially taken as the sum of contributions $\langle N_{ch}^{pp}\rangle$ from each binary collision. This naive approach, however, substantially overestimates yields. The improved descriptions employ two-component models with “soft” ($N_{part}$) and “hard” ($N_{coll}$) contributions, where multiplicity distributions are simulated via Negative Binomial Distributions fitted to experimental data, one can find more details in Ref.~\cite{KN, ALICE_centr}.

Nevertheless, SGM assumptions ignore energy–momentum conservation law and are therefore overestimates  the number of nucleon-nucleon collisions $\langle N_{coll} \rangle$. Contrary to SGM, the Modified Glauber Model (MGM) incorporates an effective energy loss in successive inelastic collisions via a single parameter $k$, the mean momentum fraction lost per nucleon-nucleon inelastic interaction. This reduces, step-by-step,  both the effective inelastic cross-section and subsequent charged particle production, providing a more realistic description of nonlinear effects in heavy-ion and light-ion collisions at RHIC and LHC.
At the LHC energies, the value of $k$ is estimated by fitting Pb+Pb and Xe+Xe multiplicity data across centralities. The MGM reproduces observed yields and significantly lowers the estimated number of binary collisions compared to SGM.

In this work, we present the MGM predictions for O+O and Ne+Ne collisions at the LHC, emphasizing the role of initial-state geometry and conservation laws. These results may be useful for future studies of QGP properties through system-size and energy-density scans.

\section{Event simulation and analysis procedure}

In this section, we will describe the generation of the nucleons configuration in  the colliding nuclei and the calculations for each nucleon-nucleon collision in our MC codes of SGM and MGM. 

\subsection{Nucleus generation}
To determine the distribution of nucleon density in the Pb, Xe, O and Ne nuclei, we will use the standard three-parameter Fermi (3pF) distribution: 
$$\rho = \frac{\rho_0 (1+\omega (\frac{z}{R})^2)}{1 + \exp{\frac{r-R}{a}}} \eqno(1)$$

The parameters for 3pF distribution were taken from Ref.~\cite{VJ, Predictions, PredictionsL} and are shown in the table:

\label{tab:professional}

\begin{table}[h!]
\centering
\caption{parameters of 3pF distribution}
\label{tab:professional}
\begin{tabular}{lccc}
\toprule
& \textbf{$a, fm$} & \textbf{$R, fm$} & \textbf{$\omega$} \\
\midrule
$ ^{208}\text{Pb}$ & 0.546 & 6.62 & 0 \\
$ ^{16}\text{O}$ & 0.513 & 2.608 & -0.051 \\
$ ^{20}\text{Ne}$ &0.698 & 2.791 & -0.168 \\
\bottomrule
\end{tabular}
\end{table}


For the $^{129}\text{Xe}$ nucleus we used \cite{LNS}:

$$\rho(x,y,z) = \rho_0 \cdot \frac{1}{1 + exp(\frac{r - R \cdot (1 + \beta_2 \cdot Y_{20} + \beta_4 \cdot Y_{40})}{a})} \eqno(2)$$
$$Y_{20} = \sqrt{\frac{5}{16\pi}}\cdot (3\cdot cos^2(\theta) - 1),~Y_{40} = \frac{3}{16\sqrt{\pi}} \cdot (35 \cdot cos^4(\theta) - 30 \cdot cos^2(\theta) + 3)$$

with $R = 5.36,~ a = 0.59,~ \beta_2 = 0.18, ~\beta_4 = 0$.

\subsection{Calculation of each nucleon-nucleon collision in Modified Glauber Model}

The key difference between the MGM presented in this work and the SGM lies in the consistent application of energy and momentum conservation laws for each binary nucleon-nucleon collision.

To account for energy loss due to  the multiple particle production, we introduce a single parameter, $k$, analogous to the approaches in Refs.~\cite{FI, FS, SF}. This parameter is universal for all nucleon-nucleon collisions and represents the average fraction of momentum retained by a nucleon in the center-of-mass (CM) frame of the colliding nucleon pair. Consequently, in any binary collision, a nucleon loses a fraction $(1-k)$ of its momentum in the CM frame.

The sequence of nucleon-nucleon collisions is determined by the distances between nucleons, meaning the closest pairs collide first. Two nucleons are considered to interact if their impact parameter $b$ satisfies the condition $b < \sqrt{\sigma_{NN}^{inel}/ \pi}$. The nucleon-nucleon cross-section $\sigma_{NN}^{inel}$ is parameterized as a function of the CM energy $\sqrt{s}$:
$$
\sigma_{NN}^{inel}(s) = a + b \cdot \ln^n(s), \eqno(3)
$$
with the parameters taken from Ref.~\cite{L}: $a = 28.84 \pm 0.52$ mb, $b = 0.05 \pm 0.02$ mb, $n = 2.37 \pm 0.12$.

Further, we move to the center-of-mass system of two nucleons and recalculate the momenta after the collision as follows:
$$
P_1' =  k \cdot P_1, \quad P_2' =  k \cdot P_2. \eqno(4)
$$

One can see a more complete description of the calculations in Ref.~\cite{SF}

For a given impact parameter, the average number of participant nucleons $\langle N_{part} \rangle$ and the average number of binary nucleon-nucleon collisions $\langle N_{coll} \rangle$ are determined by direct Monte Carlo counting during the sequence of collisions.

The average multiplicity of charged particles $\langle N_{ch}^{tot} \rangle$ is calculated by summing the contributions from each binary collision $\langle N_{ch}^{pp} \rangle$: $$\langle N^{pp}_{ch} \rangle = a + b \cdot \ln(s) + c \cdot \ln^2(s), \quad
a = 16.65,~b = -3.147,~c = 0.334 \eqno(5)$$
with parameters from Ref.~\cite{GO}. For every nucleon-nucleon collision, the average charged particle multiplicity $\langle N_{ch}^{pp} \rangle$ is determined using a parameterization Eq.~(4) as a function of the CM energy $\sqrt{s_{NN}}$ of that specific colliding nucleon pair. The total $\langle N_{ch} \rangle$ is then the sum of these contributions over all collisions in the event.

To select the parameter $k$ for O+O and Ne+Ne collisions at LHC energies, the already known, published  experimental data on Pb+Pb and Xe+Xe collisions  were used to fit  the parameter  $k$.
For our predictions of total multiplicity for light nuclei collisions  at the LHC, 
we obtained  a mean value of the parameter $k$  in the analysis of centrality dependence of charged particle multiplicity  normalized to the pair of nucleons - participants in Pb+Pb and Xe+Xe collisions at the LHC. The following data were used: in Pb+Pb collisions at $\sqrt{s_{NN}}$ =2.76 \cite{ALICEPb2760},  at $\sqrt{s_{NN}}$ =5.02 TeV\cite{ALICEPb5020} and  in Xe+Xe collisions \cite{ALICExe}  at $\sqrt{s_{NN}}$ =5.44 TeV.We show the results of fits in Figure ~\ref{fig:left}. The relevant values of $k$ are presented in  Table~\ref{tab:professional}. we  choose the mean value  $k = 0.19\pm0.05$.

\begin{table}[h!]
\label{tab:1}
\centering 
\caption{Values of parameter $k$ in fits to the data in Pb+Pb collisions at $\sqrt{s_{NN}}$ =2.76  \cite{ALICEPb2760},  5.02 TeV\cite{ALICEPb5020} and  in Xe+Xe collisions \cite{ALICExe}}
\label{tab:professional}
\begin{tabular}{lcc}
\toprule
Colliding system & $\sqrt{s_{NN}},~\text{TeV}$ & $k$ \\
\midrule
$^{208}\text{Pb}$ & $2.76$ & $0.22 ^{+0.01}_{-0.02}$ \\
\addlinespace[0.5em] 
$^{208}\text{Pb}$ & $5.02$ & $0.21^{+0.002}_{-0.01}$ \\
\addlinespace[0.5em]
$^{129}\text{Xe}$ & $5.44$ & $0.16^{+0.01}_{-0.002}$ \\
\bottomrule
\end{tabular}
\end{table}

\section{Results and discussions}

\begin{figure}[h!]
    \centering
    \begin{minipage}{0.45\textwidth}
          \centering
        \includegraphics[width=\linewidth]{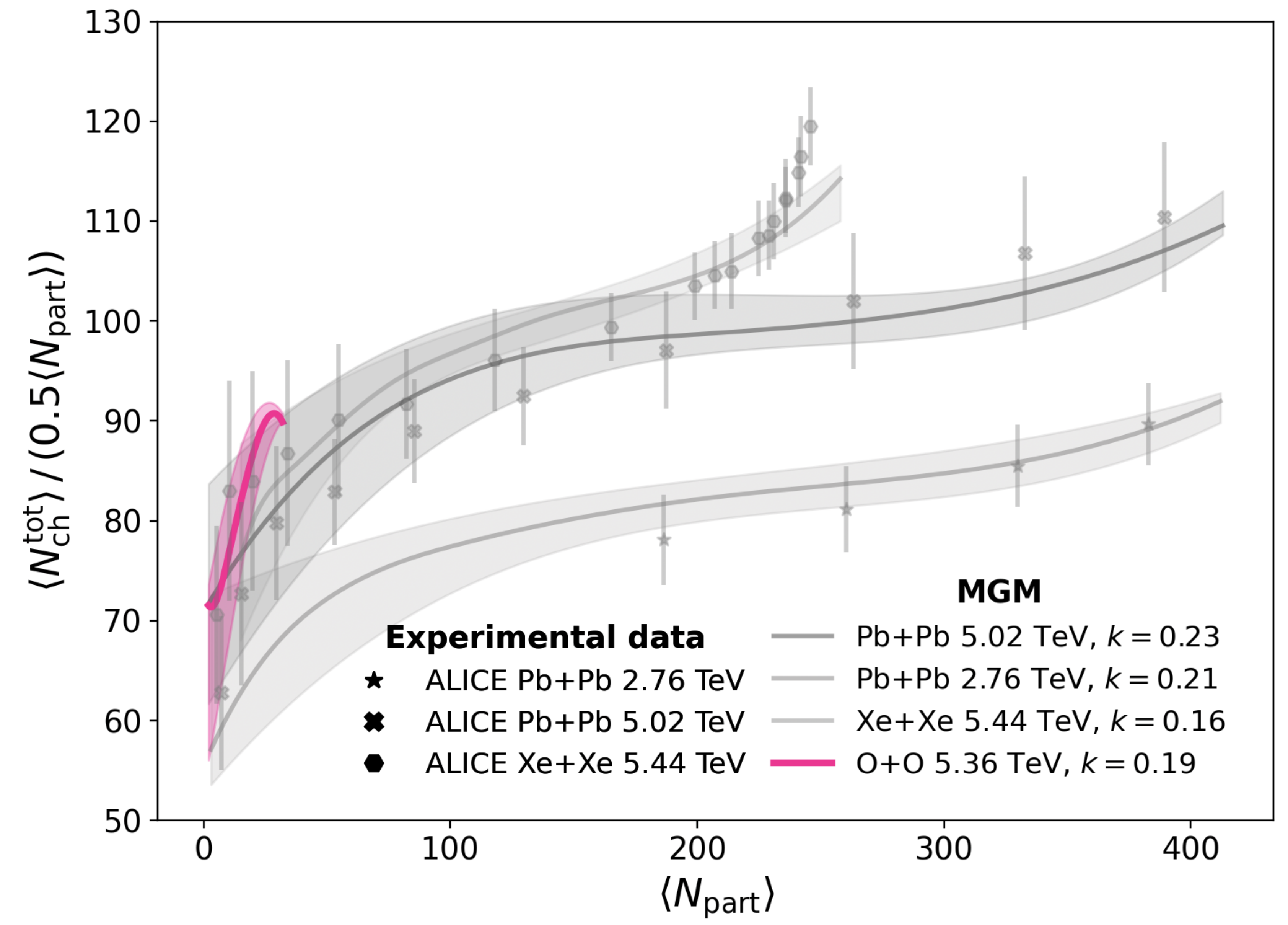}
        \caption{ Lines -- are the MGM analysis  of  the experimental data on heavy ion collisions at the LHC  for the total multiplicity normalized to the number of participating nucleon pairs. Experimental points are:  Pb+Pb collisions at $\sqrt{s_{NN}}$ =2.76 \cite{ALICEPb2760},  5.02 TeV\cite{ALICEPb5020} and   Xe+Xe collisions \cite{ALICExe}  at $\sqrt{s_{NN}}$ =5.44 TeV.
        Red line -- is the result of our prediction for O16+O16 collisions at $\sqrt{s_{NN}}$=5.36 TeV. The shaded areas are indicating the uncertainties of the MGM calculations.}
        \label{fig:left} 
    \end{minipage}
    \hfill
    \begin{minipage}{0.45\textwidth}
       \centering
        \includegraphics[width=\linewidth]{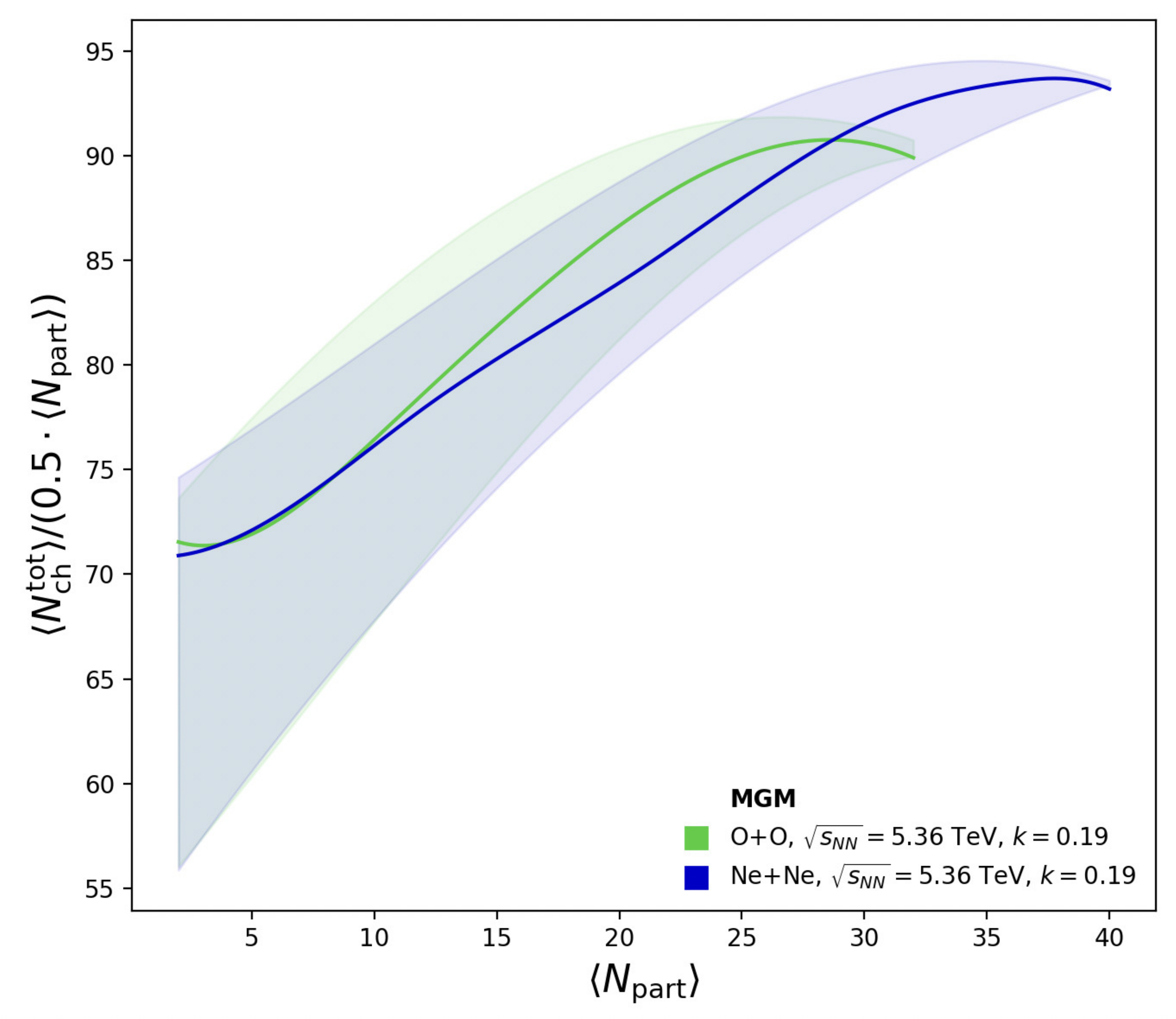}
        \caption{Predictions of the total multiplicity normalized to the number of participating nucleon pairs for O+O and Ne+Ne collisions at $\sqrt{s_{NN}}$ =5.36 TeV.
        Parameter $k = 0.19$.}
        \label{fig:right}
    \end{minipage}
\end{figure}

We demonstrate  in Figure~\ref{fig:left} our results obtained in the  MGM analysis   of  data from heavy ion collisions at the LHC:
for Pb+Pb at $\sqrt{s_{NN}}$ =2.76 \cite{ALICEPb2760},  5.02 TeV\cite{ALICEPb5020} and   for  Xe+Xe  \cite{ALICExe}  at $\sqrt{s_{NN}}$ =5.44 TeV. As we see, a single parameter Modified Glauber model is capable to describe the observed non-linear effects of total multiplicity yields in the whole region of centralities.

So, these results, obtained with  a single efficient parameter $k$ in the Modified Glauber model approach, are indicating the importance of energy-momentum account in the multiparticle production in  heavy-ion collisions at the LHC.
It should also be noted that in this case of total multiplicity yields, the role of deformation of colliding nuclei is not very important even in the region of so-called "up-tick" effects observed by ALICE for very central Pb+Pb  and  in  Xe+Xe  collisions \cite{ALICExe}. 

In Figure ~\ref{fig:left} we show also our predictions of the total multiplicity normalized to the number of participating nucleon pairs for O+O at $\sqrt{s_{NN}}$ = 5.36 TeV. Parameter $k = 0.19$. The same value of parameter $k$ is used for the prediction of  centrality dependence of yields in Ne+Ne collisions -- the results are presented in the relevant expanded region of $N_{part}$ in Figure  ~\ref{fig:right} -- along with the repeated  predictions for O+O collisions. We see that in both cases of central O+O and  Ne+Ne collisions, the  total multiplicity yields,  normalized to the number of participant nucleons, is very close to the relevant values  measured in  semi-central Pb+Pb and Xe+Xe collisions at the LHC.

The shaded areas in Fig.~\ref{fig:left} and Fig.~\ref{fig:right} show the Monte Carlo data dispersion, solid lines are the results of data splines.

\newpage

\begin{figure}[ht!]
    \centering
    \begin{minipage}{0.45\textwidth}
        \centering
        \includegraphics[width=1\linewidth]{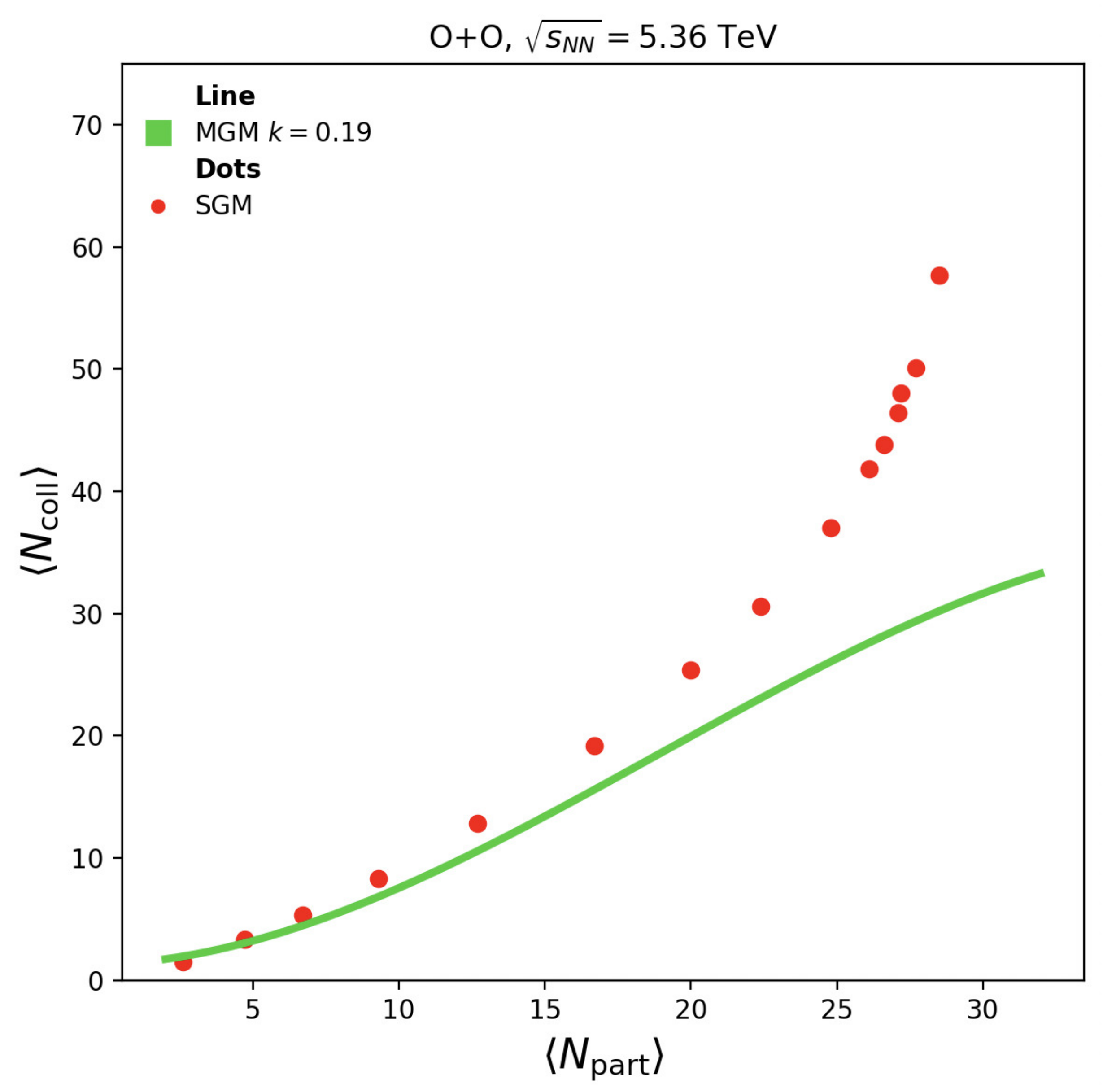}
        \subcaption{}
        \label{fig:image5}
    \end{minipage}
    \hfill
    \begin{minipage}{0.45\textwidth}
        \centering
        \includegraphics[width=1\linewidth]{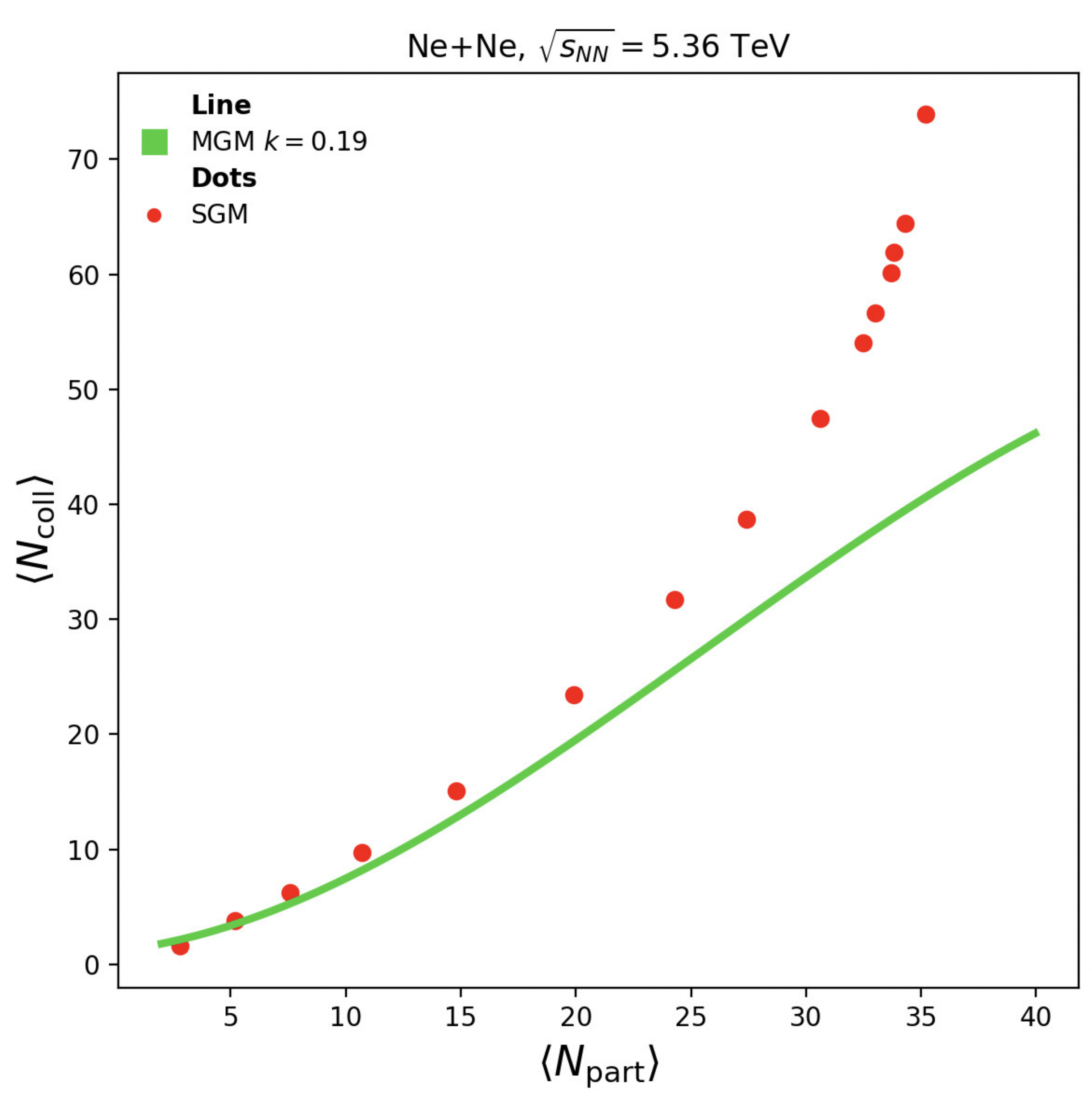}
        \subcaption{}
        \label{fig:image6}
    \end{minipage}
    
    \caption{Values of $\langle N_{coll}\rangle$ as a function of $\langle N_{part} \rangle$ at $\sqrt{s_{NN}} = 5.36~\text{TeV}$ for O+O (a) and Ne+Ne (b) collisions: comparison of MGM (green lines) to the SGM predictions (red dots). Ref.~\cite{PredictionsL}.}
    \label{fig:coll_comparison}
\end{figure}

When discussing the number of participating nucleons $\langle N_{part} \rangle$,  it should be noted that it is approximately the same in all models, as shown in Ref.~\cite{Drozhzhova}.
At the same time, comparison of the MGM results for $\langle N_{coll} \rangle$ as a function of centrality, to the SGM predictions Ref.~\cite{PredictionsL}, shows a factor of $\sim 2$ decrease both for very central O+O and Ne+Ne collisions at $\sqrt{s_{NN}} = 5.36~\text{TeV}$, see in Figure \ref{fig:image5} and in Figure \ref{fig:image6}~(green lines for the MGM).

\section{Conclusion}

We make predictions  in the framework of the Modified Glauber model for the centrality dependence  of total multiplicity yields of charged particles produced in O+O and Ne+Ne collisions at the LHC energies. The  efficient account of the energy-momentum conservation in multiparticle production in collisions of light nuclei is
taken by a single model parameter ($k$), that is  considered to be universal for all nucleon-nucleon  collisions at the LHC energies.
For predictions, we fixed the average value  $k = 0.19\pm0.05$ as defined from our analysis of data for Pb+Pb and Xe+Xe collisions at LHC. We obtain with the MGM, that  in the case of O+O and Ne+Ne collisions at  $\sqrt{s_{NN}} = 5.36~\text{TeV}$, the  total multiplicity yields,  normalized to the number of participant nucleons, is expected to  be very close to the relevant values  measured in  semi-central Pb+Pb and Xe+Xe collisions at the LHC.

We argue that the purely geometrical effects  and the energy-momentum conservation laws are driving the predictions of  the total multiplicity yields  -- both in collisions of heavy and of  light ion systems at the  LHC. The numbers of binary nucleon-nucleon collisions $\langle N_{coll} \rangle$, obtained in a single parameter  Modified Glauber model,  are found to  be remarkably different  from the Standard Glauber model approach, that neglects energy-momentum conservation laws and  overestimates  $\langle N_{coll} \rangle$.

Acknowledgments: The authors acknowledge Saint-Petersburg State University for a research project ID: 103821868 

\section{Attachment: Modified Glauber Model data}

\begin{table}[ht!]
\centering
\begin{minipage}{0.45\textwidth}
\centering
\caption{Ne+Ne at $\sqrt{s_{NN}} = 5.36~ \text{TeV}$,~$k = 0.19$}
\label{tab:b_ranges_first}
\begin{tabular}{lccc}
\toprule
$b$ range (fm) & $\langle N_{\text{part}} \rangle$ & $\langle N_{\text{coll}} \rangle$ & $\langle N_{\text{ch}} \rangle$ \\
\midrule
0.0--1.0 & 33.6 & 40.3 & 40.6 \\
\addlinespace[0.3em]
1.0--2.0 & 31.6 & 37.4 & 37.6 \\
\addlinespace[0.3em]
2.0--3.0 & 27.7 & 32.3 & 32.4 \\
\addlinespace[0.3em]
3.0--4.0 & 22.8 & 26.2 & 26.1 \\
\addlinespace[0.3em]
4.0--5.0 & 17.4 & 19.2 & 19.0 \\
\addlinespace[0.3em]
5.0--6.0 & 12.3 & 12.7 & 12.5 \\
\addlinespace[0.3em]
6.0--7.0 & 7.9 & 7.4 & 7.2 \\
\addlinespace[0.3em]
7.0--8.0 & 4.8 & 4.0 & 3.8 \\
\addlinespace[0.3em]
8.0--9.0 & 2.9 & 2.2 & 2.0 \\
\addlinespace[0.3em]
9.0--10.0 & 2.2 & 1.6 & 1.4 \\
\bottomrule
\end{tabular}
\end{minipage}
\hfill
\begin{minipage}{0.45\textwidth}
\centering
\caption{O+O at $\sqrt{s_{NN}} = 5.36~ \text{TeV}$,~$k = 0.19$}
\label{tab:b_ranges_second}
\begin{tabular}{lccc}
\toprule
$b$ range (fm) & $\langle N_{\text{part}} \rangle$ & $\langle N_{\text{coll}} \rangle$ & $\langle N_{\text{ch}} \rangle$ \\
\midrule
0.0--1.0 & 25.7 & 28.4 & 28.6 \\
\addlinespace[0.3em]
1.0--2.0 & 24.2 & 26.5 & 26.5 \\
\addlinespace[0.3em]
2.0--3.0 & 20.8 & 22.7 & 22.6 \\
\addlinespace[0.3em]
3.0--4.0 & 16.7 & 17.7 & 17.5 \\
\addlinespace[0.3em]
4.0--5.0 & 12.6 & 12.7 & 12.5 \\
\addlinespace[0.3em]
5.0--6.0 & 8.8 & 8.2 & 8.0 \\
\addlinespace[0.3em]
6.0--7.0 & 5.7 & 4.9 & 4.7 \\
\addlinespace[0.3em]
7.0--8.0 & 3.6 & 2.8 & 2.6 \\
\addlinespace[0.3em]
8.0--9.0 & 2.4 & 1.7 & 1.6 \\
\addlinespace[0.3em]
9.0--10.0 & 2.1 & 1.4 & 1.3 \\
\bottomrule
\end{tabular}
\end{minipage}
\end{table}


\newpage
\begin{thebibliography}{99}


 \bibitem{FI}
G. A. Feofilov and A. A. Ivanov, {\it J. Phys. G: Nucl. Part. Phys.} {\bf 31} (2005) 230, 
\href{https://doi.org/10.1088/0954-3899/31/5/001}{doi:10.1088/0954-3899/31/5/001}.

\bibitem{FS}
G. A. Feofilov and A. Yu. Seryakov, {\it AIP Conf. Proc.} {\bf 1701} (2016) 070001, 
\href{https://doi.org/10.1063/1.4938686}{doi:10.1063/1.4938686}.

\bibitem{SF}
S. Simak and G. A. Feofilov, {\it Phys. Part. Nucl.} {\bf 56} (2025) 877, 
\href{https://doi.org/10.1134/S1063779624702447}{doi:10.1134/S1063779624702447}.

\bibitem{KN}
D. Kharzeev and M. Nardi, {\it Phys. Lett. B} {\bf 507} (2001) 121, 
\href{https://doi.org/10.1016/S0370-2693(01)00457-9}{doi:10.1016/S0370-2693(01)00457-9}.

\bibitem{ALICE_centr}
ALICE Collaboration, {\it ALICE-PUBLIC-2018-003} (2018), 
\href{https://cds.cern.ch/record/2315401}{CERN Document Server}.

\bibitem{VJ}
H. De~Vries, C.~W. De~Jager, and C. De~Vries,  
At. Data Nucl. Data Tables \textbf{36}, 495 (1987), 
\href{https://doi.org/10.1016/0092-640X(87)90013-1}{doi:10.1016/0092-640X(87)90013-1}.

\bibitem{Predictions}
D. Behera \textit{et al.},  
Eur. Phys. J. A \textbf{58}, 23 (2022), 
\href{https://doi.org/10.1140/epja/s10050-022-00823-6}{doi:10.1140/epja/s10050-022-00823-6}.

\bibitem{PredictionsL}
C. Loizides, \href{https://arxiv.org/abs/2507.05853}{arXiv:2507.05853}.

\bibitem{LNS}
C. Loizides, J. Nagle, and P. Steinberg,  
\href{https://arxiv.org/abs/1408.2549}{arXiv:1408.2549 [nucl-ex]}.

\bibitem{L}
C. Loizides,  
Phys. Rev. C \textbf{94}, 024914 (2016), 
\href{https://doi.org/10.1103/PhysRevC.94.024914}{doi:10.1103/PhysRevC.94.024914}.

\bibitem{GO}
J. F. Grosse-Oetringhaus and K. Reygers, {\it J. Phys. G: Nucl. Part. Phys.} {\bf 37} (2010) 083001, 
\href{https://doi.org/10.1088/0954-3899/37/8/083001}{doi:10.1088/0954-3899/37/8/083001}.


\bibitem{ALICEPb2760}
ALICE Collaboration, Phys. Lett. B \textbf{726}, 610 (2013), 
\href{https://doi.org/10.1016/j.physletb.2013.09.022}{doi:10.1016/j.physletb.2013.09.022},  
\href{https://arxiv.org/abs/1304.0347}{arXiv:1304.0347}.

\bibitem{ALICEPb5020}
ALICE Collaboration, Phys. Lett. B \textbf{772}, 567 (2017), 
\href{https://doi.org/10.1016/j.physletb.2017.07.017}{doi:10.1016/j.physletb.2017.07.017},  
\href{https://arxiv.org/abs/1612.08966}{arXiv:1612.08966}.

\bibitem{ALICExe}
ALICE Collaboration, \href{https://arxiv.org/abs/1805.04432}{arXiv:1805.04432 [nucl-ex]}.

\bibitem{Drozhzhova}
T. A. Drozhzhova, V. N. Kovalenko, A. Yu. Seryakov and G. A. Feofilov, {\it Phys. At. Nucl.} {\bf 79} (2016) 737, 
\href{https://doi.org/10.1134/S1063778816040074}{doi:10.1134/S1063778816040074}.


\end{thebibliography}
\end{document}